\documentclass{aastex}
\shortauthors{Alpar}
\shorttitle{On Young Neutron Stars}
\begin{document}
\title{ON YOUNG NEUTRON STARS AS 
PROPELLERS AND ACCRETORS WITH CONVENTIONAL MAGNETIC FIELDS}
\vspace{3cm}
\author{M. Ali Alpar} 
\affil{Faculty of Engineering and Natural Sciences, Sabanc{\i} University,
Orhanl{\i}-Tuzla, Istanbul 81474, Turkey}


\begin{abstract}
The similarity of rotation periods of, 
the anomalous X-ray pulsars (AXPs), the soft gamma ray
repeaters (SGRs) and the dim isolated thermal neutron stars (DTNs) suggests
a common mechanism with an asymptotic spindown phase, extending through the
propeller and early accretion stages. The DTNs are interpreted as sources in
the propeller stage. Their luminosities arise from frictional heating in
the neutron star. 
If the 8.4 s rotation period of the DTN RXJ 0720.4-3125 is close to
its rotational equilibrium period, the estimated propeller torque indicates
a magnetic field in the 10$^{12}$ Gauss range. The mass inflow rate onto the
propeller is of the order of the accretion rates of the AXPs.
The limited range of rotation periods, taken to be close to equilibrium 
periods, and conventional magnetic fields in the range 
5 $\times$ 10$^{11}$ - 5 $\times$ 10$^{12}$ Gauss correspond to a 
range of mass inflow rates 3.2 $\times$ 10$^{14}$ gm s$^{-1} < \dot{M} 
<$ 4.2 $\times$ 10$^{17}$ gm s$^{-1}$. Observed
spindown rates of the AXPs and SGRs also fit in with estimates
for these magnetic fields and equilibrium periods. 
The source of the mass inflow is a remnant accretion
disk formed as part of the fallback during the
supernova explosion. These classes of sources thus represent the
alternative pathways for those neutron stars that do not become radio
pulsars. For the highest mass inflow rates the propeller action may support 
enough circumstellar material so that the optical thickness to electron 
scattering destroys the X-ray beaming, and the rotation period is not 
observable.
These are the radio quiet neutron stars (RQNSs) at the 
centers of supernova remnants 
Cas A, Puppis A, RCW 103  and  296.5+10. The statistics and ages of
DTNs suggest that sources in the propeller phase are quite common, 
maybe accounting for the majority of neutron stars formed in supernovae. 
AXPs are the rare cases whose $\dot{M}$ history allows them to reach the 
post-propeller accretion phase. This model obviates the need to postulate 
magnetars for AXPs and DTNs. Frequently sampled timing 
observations of AXPs, SGRs 
and DTNs can distinguish between this explanation and the magnetar model. 
\end{abstract}
\section{Introduction}

Anomalous x-ray pulsars (AXPs) are characterized by periods 
in the range 6-12 secs (see Mereghetti 1999 for a review and 
an extensive list of references therein for properties of specific AXPs). 
RXJ0720.4-3125, one of the
dim nearby ROSAT point sources thought to be thermally
emitting neutron stars (Walter, Wolk \& Neuhauser 1996, 
Walter \& Matthews 1997) has a measured period (Haberl et al. 1997) of 8.4 s,
while a recently discovered member of this class, RXJ0420.0-5022, 
is reported to have a period of 22.7 s (Haberl, Pietsch \& Motch 1999).  
Among the four confirmed soft gamma ray repeaters 
(SGRs- Mazets et al. 1979, 1981, Cline et al. 1980, Woods et al. 1999) with 
quiescent X-ray sources, there are two with measured pulse periods. 
Both periods are in the AXP period range, with P = 7.47s for SGR 1806-20 
(Kouveliotou et al. 1998) and P = 5.16 s for SGR 1900+14 (Hurley et al.1999, 
Kouveliotou et al. 1999). For the AXPs it was pointed 
out (van Paradijs, Taam \& van der Heuvel 1995, Ghosh, Angelini \& White 1995) 
that these periods would obtain as equilibrium periods for the accretion 
rates $\dot{M}\sim 10^{15}gm/s$ corresponding to the observed luminosities 
and for 10$^{11}$-10$^{12}$ G magnetic fields. 

Values of $\dot{P}$ measured from the AXPs and SGRs, together with the long
rotation periods in the 5-12 s range have led to the suggestion that these
sources are magnetars (Kouveliotou et al. 1998, 1999, 
Duncan \& Thompson 1992, Thompson \& Duncan 1993, 1995), 
neutron stars with very strong magnetic
fields B$\sim $ 10$^{14}$-10$^{15}$ G 
spinning down through magnetic dipole radiation.
The short cooling age estimated for RXJ0720.4-3125 has also been 
suggested (Heyl \& Hernquist 1998) as evidence for a 
magnetar if the source is an isolated pulsar born 
with a short rotation period. We propose here 
an alternative explanation, extending the accretion hypothesis to
include propeller phases, and linking these classes of neutron stars 
as stars with conventional $\sim$ 10$^{12}$ G magnetic fields. Mass 
inflow at rates $\dot{M} \sim$ 5 $\times$ 10$^{15}$ - 4 $\times$ 
10$^{16}$ gm/s from a disk (van Paradijs, Taam \& van den Heuvel 1995, 
Ghosh, Angelini \& White 1997) around the neutron 
star will produce torques on the neutron star of the order of the spindown 
torques observed. The AXPs and SGRs are accreting at least part of the 
incoming mass flow. The similarity of the periods simply reflects similar 
conditions. All these systems are
asymptotically approaching equilibrium periods in a common range, defined by
the common ranges of magnetic fields and mass transfer rates. As the
approach to rotational equilibrium is asymptotic, $P/\dot{P}$ is not the
true age of these systems. Sources of different ages and different
circumstances with similar conventional magnetic fields have similar periods 
in this asymptotic regime under a fairly large range of mass
inflow rates. 

We propose that the DTNs, AXPs, SGRs and RQNSs constitute the 
alternative subclasses of young neutron stars produced by supernovae, 
complementary to the subclass of neutron stars that show up as radio 
pulsars (Kaspi 1999). 
In view of the plausible associations of several AXPs and SGRs 
with supernova remnants, tight upper limits on the mass of a 
binary companion (Mereghetti, Israel \& Stella 1998, Wilson et al. 1998), 
and statistics and age considerations,  
these sources are likely to be young neutron stars under mass inflow 
from a remnant disk formed by fallback material from the supernova explosion 
(Chevalier, 1989, Lin, Woosley \& Bodenheimer 1991, Mineshige, 
Nomoto \& Shigeyama 1993). 

Most ideas discussed in this paper and their links to related subjects were 
first presented in an earlier preprint (Alpar 1999). The DTNs are the 
propeller counterparts to the accreting AXPs. 
Their luminosity is due to energy dissipation in the neutron star. 
They may well account for a large fraction of neutron stars formed in 
supernovae. 
For the AXPs the scenario of accretion from a remnant disk, preceded with a 
propeller phase, has been explored also by 
Chatterjee, Hernquist \& Narayan (1999), who have employed specific thin 
disk evolution models for the time dependence of the mass inflow from the 
remnant disk, while Marsden et al. (1999) have evaluated the 
supernova associations of AXPs and SGRs to argue that what unites 
these sources is their atypical SNR environments 
rather than an intrinsic property of the neutron stars 
like superstrong magnetic fields. Under the working hypothesis of the present 
work, that all supernovae form neutron stars with 
conventional $\sim$ 10$^{12}$ G magnetic fields, the differences between the 
DTNs, AXPs and SGRs are due to their different mass inflow 
environments and histories. At the highest mass inflow rates we 
have the RQNSs. These are propellers whose 
pulse periods are not observable because of accumulated circumstellar material 
that is optically thick to electron scattering. The subclasses unified 
in this scenario represent the entire range of mass inflow rates, extending 
from $\dot{M}\sim 0$ for the radio pulsars to near 
Eddington rates inferred for the RQNSs. 

Table 1  displays observed 
parameters and model estimates for the AXPs, for the two SGRs with $\dot{P}$ 
measurements,  the two DTNs with 
measured periods, and for a typical RQNS. Section 2 discusses the energy 
dissipation in a neutron star under an external torque. Section 3 introduces 
propeller torques and applies these ideas to DTNs. Section 4 presents a 
simple model for asymptotic spindown. Section 5 relates the different 
classes of neutron stars, including the RQNSs, as supernova products 
under mass inflow. Section 6 reviews the expected signature of a remnant disk 
around the neutron star, and Section 7 presents the conclusions.\\ 

\section{Luminosity of a Non-accreting Neutron Star from Energy Dissipation}

The DTNs yield fits to blackbody spectra with temperatures of 57 eV for
RXJ185635-3754 and 79 eV for RXJ0720.4-3125 and luminosities in 
the $L_x \sim 10^{31-32}$ erg s$^{-1}$ range (Walter, Wolk \& Neuhauser 1996, 
Walter \& Matthews 1997, Haberl et al. 1997, Motch \& Haberl 1998, 
Kulkarni \& van Kerkwijk 1998). Several further candidates for
this class (Stocke et al. 1995, Haberl, Motch \& Pietsch 1998, 
Schwope et al. 1999, Motch et al. 1999, 
Haberl, Pietsch \& Motch 1999) also have similar
blackbody temperatures, flux values and limits on the ratio of X-ray flux to
optical flux. Accretion from the interstellar medium would require
unlikely ambient interstellar medium densities and low 
velocities (for a discussion of accretion from the ISM for the DTNs 
see Treves et al. 1999 and references therein). The
presence of several dim thermal neutron stars within $\sim$ 100 pc 
suggests that
there are $\sim$ 10$^4$ such sources in the galaxy, requiring ages of $\sim $
10$^6$ years or longer, if the birth rate is 10$^{-2}$ yr$^{-1}$ or less.
The cooling of a young neutron star can typically yield the observed thermal
luminosities at ages of the order of 10$^5$-10$^6$ yrs (\"Ogelman 1995). For a
neutron star born with a rotation period of the order of 10 ms, (as
typically inferred from the P and $\dot{P}$ values of young pulsars), to
have spun down to the 8.4 s period of RXJ0720.4-3125 in 10$^5$ yrs as a
rotating magnetic dipole would require a mean spindown rate $\dot\Omega$ of
the order of 10$^{-13}$ rad s$^{-2}$, and a magnetic field of the order 
of 10$^{14}$ G or more. (Heyl \& Hernquist 1998).

There is an alternative source of the thermal luminosity which takes over at 
$\sim$ 10$^5$-10$^6$ yrs, after the initial cooling, and lasts longer than
the cooling luminosity: There will be energy dissipation (frictional
heating) in a neutron star being spun down by some external torque. The rate
of energy dissipation is given by (Alpar et al. 1984, 
Alpar, Nandkumar \& Pines 1985)
\begin{equation}
\dot{E}_{diss} = I_p \omega |{\dot\Omega}|
\end{equation}
where I$_p$ is the moment of inertia of some component of the neutron star
whose rotation rate is faster than that of the observed crust by 
the amount $\omega$. $\dot{E}_{diss}$ will supply the thermal luminosity of a
non-accreting neutron star at ages greater than $\sim$ 10$^6$ 
years (Umeda et al. 1993), as the cooling luminosity rapidly falls 
below $\dot{E}_{diss}$ after the transition 
from neutrino cooling to surface photon 
cooling. Among the radio
pulsars with X-ray emission (\"Ogelman 1995, Becker \& Tr\"umper 1997), 
observations (Alpar et al. 1987, Yancopoulos, Hamilton \& Helfand 1994) 
of the pulsar
PSR 1929+10, whose spin-down age is 3.1 $\times$ 10$^6$ yrs, provide an
upper limit to the thermal luminosity which 
yields I$_p \omega <$ 10$ ^{43}$ gm cm$^2$ rad s$^{-1}$ (Alpar 1998).

A lower limit to $\dot{E}_{diss}$ can be obtained from the parameters of
large pulsar glitches (Alpar 1998). The consistency of the observed glitch
parameters of all pulsars with large glitches 
($\Delta\Omega/\Omega \geq$ 10$^{-7}$) and measured second 
derivatives $\ddot{\Omega}$ of the rotation rate (Shemar \& Lyne 1996) 
as well as the statistics 
of the large glitches  (Alpar \& Baykal 1994) support the
hypothesis that large glitches are a universal feature of pulsar dynamics.
The current phenomenological description (Alpar et al. 1993, Alpar 1998) 
of large glitches entails
angular momentum exchange between the crust and an interior component (a
pinned superfluid in current models) which rotates faster than the crust by
the lag $\omega $. The typical glitch related change 
in relative rotation rate of the crust and 
interior, $\delta \omega \sim $ 10$^{-2}$ rad s$^{-2}$, inferred 
from the common behaviour of all pulsars with large glitches,
must be less than the lag $\omega $. Using values of I$_{p}\sim $ 10$^{43}$
gm cm$^{2}$ inferred from the detailed postglitch timing measurements
available for the Vela pulsar glitches (Alpar et al. 1993) we obtain the lower 
bound I$ _{p}\omega >$I$_{p}\delta \omega $=10$^{41}$ gm cm$^{2}$ rad s$^{-1}$. This
lower bound is independent of the details of the glitch models and rests
only on the assumption that the large glitches involve angular momentum
exchange within the neutron star. Further, as the mode of angular momentum
transfer inside the star depends only on neutron star structure, the same
parameter I$_{p}\omega$ would determine the energy dissipation rates in all
neutron stars under external torques, also when the source of the external
torque is not magnetic dipole radiation. The upper and lower bounds 
together imply 
\begin{equation}
10^{41}|\dot{\Omega}| erg s^{-1} < L=\dot{E}_{diss}<10^{43}|\dot{\Omega}| 
erg s^{-1}.
\end{equation}
The luminosities of RXJ0720.4-3125, RXJ185635-3754 and the other DTNs 
(Treves et al. 2000) give  
\begin{equation}
|\dot{\Omega}|\sim 10^{-13}-10^{-10} rad s^{-2}.
\end{equation}
The measured spindown rates of AXPs and SGRs all fall in this range,
providing another similarity, in addition to the periods, between the
DTNs, AXPs and SGRs. AXP spindown rates actually suggest that
the DTN spindown rate may be closer to $|\dot{\Omega}|\sim$
10$^{-12}$ rad s$^{-2}$, which means that $\dot{E}_{diss}$ is
closer to the upper limit in Eq.(2). 
With the 8.4s period of RXJ0720.4-3125 these spindown rates imply surface
magnetic fields in excess of 10$^{14}$ G if magnetic dipole spindown is
assumed. Are there other spindown mechanisms that will give high spindown
rates, larger than 10$^{-12}$ rad s$^{-2}$, with 10$^{12}$ G magnetic fields
typical of the canonical radio pulsars and of the accreting neutron stars
with observed cyclotron lines?\\

\section{Propeller Spindown}

High spindown rates, larger than 10$^{-12}$ rad s$^{-2}$, can indeed be
expected for neutron stars with conventional 10$^{12}$ G fields 
under the typical spindown torques for certain phases of accreting sources.
For the AXPs accretion is a possibility that has already been 
explored (Mereghetti \& Stella 1995, van Paradijs, 
Taam \& van den Heuvel 1995, Ghosh, Angelini \& White 1995) 
and will be pursued below. In connection with the DTNs, we note
that a neutron star subject to mass inflow will experience high spindown
rates even when the inflowing mass is not accreted, because of the star's
centrifugal barrier (the propeller effect) (Illarionov \& Sunyaev 1975). 
We propose that RXJ
0720.4-3125 and the other DTNs are neutron stars with magnetic fields of the
order of 10$^{12}$ G, spinning down under propeller torques. There is no
accretion yet in the propeller phase. The luminosities of the DTNs are
produced by energy dissipation in the neutron star. These luminosities are
dimmer, by several orders of magnitude, than the accretion luminosities that
would have been produced if the mass inflow were accreted. The propeller torques
depend on the magnetic moment of the neutron star and on the rate of mass
inflow. Using the spindown rates estimated from the energy dissipation
luminosities together with order of magnitude expressions for propeller
torques, we show that the neutron star has a dipole magnetic field 
of the order of 10$^{12}$ G. The estimated range of mass inflow rates
coincides with the range of mass accretion rates inferred from the X-ray
luminosities of the accreting AXPs and SGRs, suggesting that the DTNs are in
the propeller stage and the AXPs and SGRs in the accretion phase under
similar mass inflow circumstances.

Occasional quiescent states of some X-ray transients are likely examples
of the propeller phase (Cui et al 1997, Zhang et al 1998, 
Campana et al 1998, Menou et al. 1999).
Sources that are persistently in the propeller phase have not been detected 
previously. This is understandable since they are not lit up with an accretion 
luminosity. Thus RXJ 0720.4-3125, RXJ185635-3754 and other DTNs 
are the first observed examples of propellers, observed only 
through their dissipation luminosities, only as
the nearest members of the DTN class, and only by ROSAT, in view of their
surface temperatures in the soft X-ray band.

In the propeller phase there is a spindown torque exerted on the neutron
star, through the interaction of its magnetosphere with the ambient
material. The order of magnitude of this spindown torque is 
\begin{equation}
N = I|\dot{\Omega}| = \mu ^{2}/{r_{A}}^{3}
\end{equation}
where $\mu =BR^{3}$ is the magnetic moment of the neutron star with surface
dipole magnetic field B and radius R, and 
\begin{equation}
r_{A}=9.85\times 10^{8}cm {\mu _{30}}^{4/7}{{\dot{M}}_{15}}^{-2/7} m^{-1/7}
\end{equation}
is the Alfven radius (Lipunov 1992, 
Frank, King \& Raine 1992). 
Here $\mu _{30}$ denotes the magnetic moment
in units of 10$^{30}$ G cm$^{3}$, ${\dot{M}}_{15}$ is the mass inflow rate
in units of 10$^{15}$ gm s$^{-1}$ and m is the neutron star mass in solar
mass units. The propeller phase will last until accretion starts when a
critical rotation period is reached. This critical period is of the order
of, but somewhat smaller than the equilibrium rotation period 
\begin{equation}
P_{eq}=16.8 s{\mu _{30}}^{6/7}{{\dot{M}}_{15}}^{-3/7} m^{-5/7}
\end{equation}
which obtains when the star's rotation rate equals the Keplerean rotation
rate of ambient matter at the Alfven radius; that is, when the corotation
radius r$_{c}$ = (GM)$^{1/3}\Omega ^{-2/3}$ becomes equal to r$_{A}$ . 
In the propeller phase the source settles to an asymptotic 
spindown towards the equilibrium. A source
is most likely to be observed during its asymptotic phase. Since in the
asymptotic regime r$_{c}$ is close to r$_{A}$ we can estimate the magnetic
moment of a propeller from its spindown rate, and rotation rate, without
knowing the mass inflow rate, simply by substituting r$_{c}$ in Eq.(4). This
leads to
\begin{equation}
\mu =\frac{(I|\dot{\Omega}|GM)^{1/2}}{\Omega }. 
\end{equation}
We take the DTN RXJ0720.4-3125, with P = 8.4 s, as our typical 
example (Haberl et al. 1997). 
The other DTN with a reported rotation 
period, RXJ 0420.0-5022, P = 22.7 s (Haberl, Pietsch \& Motch 1999), yields 
similar results. 
For RXJ0720.4-3125 we obtain $\mu _{30}\sim (0.35-3.5)$ 
using the bounds on the spindown rate, Eq.(2). Thus if RXJ0720.4-3125 is a
propeller it has a conventional magnetic field, 
\begin{equation}
3.5 \times 10^{11} G < B < 3.5 \times 10^{12} G
\end{equation}
Using Eq.(6) and setting the observed period to be equal to the 
equilibrium period, we can now obtain bounds on the mass inflow rates: 
\begin{equation}
{{\dot{M}}_{15}} = (16.8 s/P_{eq})^{7/3}({\mu _{30})}^{2}m^{-5/3}. 
\end{equation}
This yields ${{\dot{M}}_{15}}\sim (0.62-62)$ for the mass inflow rate for
RXJ0720.4-3125. Model parameters obtained from Eqs. (7) and (9) are given 
in Table 1, assuming $|\dot{\Omega}| = $ 2.6 $\times$ 10$^{-12}$ rad s$^{-2}$ 
from Eq.(2) for RXJ0720.4-3125, and $|\dot{\Omega}| = $ 10$^{-11}$ 
rad s$^{-2}$ for RXJ 0420.0-5022. 

For most DTNs and for all RQNSs, which, as we 
propose below, are also propeller candidates, we do not have observed 
rotation periods. For these sources, setting the rotation period to be equal to the equilibrium period, 
\begin{equation}
{{\dot{M}}_{15}} \cong {\mu _{30}}^{-1/3}{|\dot{\Omega}|_{-12}}^{7/6}
m^{-1/2} {I_{45}}^{7/6}
\end{equation}
from Eqs.(6) and (7). With the range of $|\dot{\Omega}|$ obtained from 
DTN luminosities (Eqn. (3)), ${{\dot{M}}_{15}}\sim (0.1-200)$ 
is obtained for these DTNs for $\mu _{30}$ = 1.  

If the source of mass inflow is not depleted, the propeller phase is 
followed by an accretion phase starting at some critical 
period. The neutron star will continue to spin down, now as an 
accreting source, as its period evolves from the critical period 
towards the equilibrium period.
Similarity of the rotation periods of the AXPs with the 8.4 s period of
the DTN RXJ0720.4-3125 suggests that the approach to rotational equilibrium 
is an asymptotic process extending through the propeller phase on 
to the accretion phase. These sources are asymptotically close to their 
individual equilibrium rotation periods, which lie in a narrow range 
determined by their magnetic moments and mass inflow histories.

The mass accretion rates inferred from the X-ray luminosities of the AXPs
and SGRs all fall in the range ${{\dot{M}}_{15}}\sim 1-100$. 
The corresponding magnetic moments for
accreting sources near rotational equilibrium are in the
10$^{12}$ G range given the 5-15 s rotation periods: 
\begin{equation}
{\mu_{30}} = (P_{eq}/16.8 s)^{7/6} {{\dot{M}}_{15}}^{1/2} m^{5/6}.
\end{equation}

To summarize, (i) the spindown rates inferred for the DTN sources through
the interpretation of their luminosities as due to energy dissipation
(Eqs.(1)-(3)) agree with the observed spindown rates for the AXPs and SGRs
to order of magnitude. (ii) Estimates of the magnetic moment, from the
spindown rates for the DTNs as propellers close to rotational equilibrium,
coincide in the same range as the magnetic moments inferred from the mass
accretion rates implied by the X-ray luminosities of AXPs and SGRs as accretors
asymptotically close to rotational equilibrium. This is the 
conventional 10$^{12}$ G range of most young neutron stars observed as 
rotation powered pulsars and in high mass X-ray binaries. 
(iii) Furthermore, the mass inflow
rates inferred for the DTNs from the assumption that they are close to
rotational equilibrium, agree with the accretion rates of the AXPs and SGRs.
These observations strongly support the unifying hypothesis that the sources
are related instances of asymptotic spindown near 
similar equilibrium periods.

\section{The Asymptotic Spindown Regime}

We now turn to a simple model for the asymptotic spindown. A neutron star in
the presence of inflowing matter experiences both spindown and spin-up
torques. The long term evolution is determined by the balance between these,
described by a function $n(\Omega/\Omega_{eq} - \omega_c)$ 
(Ghosh \& Lamb 1991) which
goes through zero when $\Omega = \omega_c \Omega_{eq} $ where $\omega_c$ is
of order one, and $\Omega_{eq} = 2\pi /P_{eq}$. In accretion from a
disk, the relative specific angular momentum brought in by the accreting
material to spin the neutron star up is $[(GMr_A)^{1/2} - \Omega {r_A}^2]$.
Since the dimensional torque is $\mu^2 / {r_A}^3 \sim \dot{M} (GMr_A)^{1/2}$, 
spindown near equilibrium can be modelled as 
\begin{equation}
\dot\Omega = (\mu^2 / I {r_A}^3) ( 1 - \Omega/\Omega_{eq}) = ( \Omega_{eq} -
\Omega ) / t_0
\end{equation}
where t$_0$ = $\Omega_{eq} I {r_A}^3 / \mu^2 $. Here the zero of the torque
(the end of the spindown era) is taken to be at $\Omega = \Omega_{eq}$
rather than $\omega_c \Omega_{eq} $, for simplicity. The AXPs and SGRs must
then be sources that have emerged from propeller spindown and started to accrete
while still under a net spindown torque as they have not yet reached rotational
equilibrium. Another simplifying assumption we make is that the mass inflow rate
is constant in each source. 
In a more realistic model the mass inflow rate should
be time dependent, especially if its source is a remnant disk with a
finite lifetime, rather than a binary companion. A decaying $\dot{M}$ leads to a
natural explanation for the prevalence of spindown in an accreting source: the
equilibrium period increases as $\dot{M}$ decreases, and the source will be
spinning down as its period tracks the 
equilibrium period (Ghosh, Angelini \& White 1997, 
Chatterjee, Hernquist \& Narayan 1999), answering the criticism of 
accretion models raised by Xi (1999). Chatterjee, Hernquist \& Narayan 
(1999) have
employed an $\dot{M}$ decaying as a power law in time, corresponding to the
viscous evolution of a thin  disk (Mineshige, Nomoto \& Shigeyama 1993, Canizzo et al 1997).
The actual situation warrants taking into account propeller boundary 
conditions. The propeller activity could probably support a thick disk or 
corona. It would also lead to a different characteristic time 
dependence of $\dot{M}$. Here we take a constant $\dot{M}$ 
to represent an average of the actual $\dot{M}$(t).

Even under constant mass inflow rate, the 
torque history must be more complicated.
In X-ray binaries, the occurrence of spindown and spinup torques of comparable 
magnitudes, and torque reversals on all timescales signals the 
importance of electromagnetic torques. The real torques could lead to initial
power law decays of $\Omega$ as a function of time. 
Whatever the form of the initial spindown may be, once the
rotation rate is close to the equilibrium value, the 
factor $(\Omega_{eq} - \Omega )$ dominates the asymptotic evolution, which 
becomes an exponential decay. Here we employ the simple 
asymptotic model of Eq.(12) assuming that the magnetic moment is constant, and 
that the mass inflow rate is also a constant representing the long term average 
$\dot{M}$. The results are given in Table 1 and Figure 1. 
With constant $\mu$ and $r_A$ , the spindown leads $\Omega$ 
towards $\Omega_{eq}$: 
\begin{equation}
\Omega (t) = [\Omega (0) -\Omega_{eq}] exp ( -t / t_0) + \Omega_{eq}.
\end{equation}
This solution is to describe the spindown through both the propeller and the
accretion phases, with mass accretion starting at 
some $\Omega_{acc} < \Omega_{eq}$. From Eqs.(5),(6) and (12) we obtain 
\begin{equation}
t_0 = 1.3 \times 10^{12} s I_{45} {{\dot{M}}_{15}}^{-1} m^{-2/3} {\Omega_{eq}
}^{4/3} .
\end{equation}
Substituting the expression for t$_0$ in the spindown equation, 
Eq.(12), $\Omega_{eq}$ can be obtained for each source with known $\Omega$ 
and $\dot\Omega$ by solving 
\begin{equation}
\Omega_{eq} = \Omega + 1.3 \times 10^{12} s \dot\Omega I_{45} {{\dot{M}}_{15}
}^{-1} m^{-2/3} {\Omega_{eq}}^{4/3} .
\end{equation}
Once the equilibrium rotation rate is estimated, the magnetic moment $\mu$
can be obtained from Eq.(11). 
The time since the beginning of the propeller phase can be obtained from
Eq.(13) for a given initial rotation period, which can be taken to 
be P(0) = 0.01 s for a newborn neutron star. 
The age t estimated from Eq.(13) is: 
\begin{equation}
t = t_0 log [(\Omega(0) - \Omega_{eq})/ (\Omega - \Omega_{eq})].
\end{equation}
Sample solutions are given in Table 1 for the AXPs, the 
SGRs with measured P and $\dot{P}$ and for the DTNs with measured 
periods. The AXPs show changes in spindown rate, by up to an order of
magnitude, on timescales of several years. Long term average values of the
spindown rate are used, as appropriate for the model to describe the long
term average trend. Mass inflow rates are inferred from the observed
luminosities taken as accretion luminosities onto neutron stars, 
$\dot{M_{acc}}$ = RL/GM . The luminosity has been observed to 
change by as much as a factor of 15 in AXPs (Torii et al. 1998). 
Furthermore, the mass inflow rate from a remnant disk will be 
decreasing as a function of time, so that the constant 
representative $\dot{M}$ appropriate for our simple model should be 
larger than the present time $\dot{M_{acc}}$ inferred from the acretion 
luminosity. For 1E1048.1-5937, the current $\dot{M_{acc}}$ 
leads to P$_{eq}$ = 64 s., while for 1E2259+586 the current $\dot{M_{acc}}$ 
gives a relaxation time t$_0$ that may be larger than the age of the associated SNR. For these two sources, taking examples of mass inflow rates 
$\dot{M} > \dot{M_{acc}}$, one can obtain solutions in agreement 
with the asymptotic model. These solutions are given in Table 1.

Three out of six AXPs provide direct evidence of youth through their likely
supernova associations. These AXPs are situated close to the centers of
the SNRs. The associations with SNRs are not as certain for the SGRs:
The SGRs are at edges of the SNRs, requiring velocities $\sim$ 1000 Km
s$^{-1}$ if they were born at the center (for a recent discussion of the 
SNR associations of the AXPs and SGRs, and references to the 
individual associations, see Gaensler 1999). For both AXPs and SGRs, the
associated SNRs are not typical; but rather of a particularly dense
morphology. This point was realized and elaborated by Marsden et al
(1999) in their discussion of propeller and accretion scenarios, as
evidence that these classes of sources are related through similarities
in their environments rather than a common intrinsic property like very
strong magnetic fields. Here we adopt the possibility that a remnant
disk around the neutron star, formed from the debris of the core
collapse in a supernova explosion (Chevalier 1989, Mineshige, 
Nomoto \& Shigeyama 1993), provides the mass inflow. 
The propeller and accretion phases
would terminate when the supply of $\dot{M}$ is depleted.

\section{The Fallback Mass Inflow and the Signature of the 
Neutron Star Born in a Supernova}

Does the narrow range of observed rotation periods indicate a very
restricted range of $\dot{M}$ ? The range of estimates for the sources in 
Table 1 extends from 10$^{15}$ gm s$^{-1}$ to 10$^{17}$ gm s$^{-1}$. 
To estimate the parameter range corresponding to the narrow range of 
equilibrium periods, we note that the magnetic moment and mass inflow 
rate or equivalently, the magnetic moment and equilibrium rotation rate 
are independent parameters. For magnetic moments in the 
range 5 $\times$ 10$^{11}$ G - 5 $\times$ 10$^{12}$ G, the range of mass 
inflow or accretion rates which would lead to equilibrium periods 
of 5-15 s extends over three orders of magnitude, up to about half the
Eddington rate: 
\begin{equation}
3.2 \times 10^{14} gm s^{-1} < \dot{M} < 4.2 \times 10^{17} gm s^{-1}.
\end{equation}
This is not a very restricted range. Fig.1 shows the mass inflow 
rate as a function of the equilibrium period for 0.5 $ < \mu_{30} < $ 5. 
Also shown are the model solutions for equilibrium periods and mass 
inflow rates for all the sources presented in Table 1. The narrow range of 
observed periods lies between the vertical lines. We note that the narrow range of observed and equilibrium periods corresponds to a rather wide range of 
$\dot{M}$, extending over more than two orders of magnitude. Sample solutions 
for different sources are scattered in $\dot{M}$ and $\mu$, as they 
should be, since the magnetic moments of neutron stars should not be correlated with the mass inflow from their environments. This scatter indicates that 
the model does not require an artificial correlation between these parameters. 

The mass inflow rates we obtain are all above $\sim$ 10$^{14}$ gm s$^{-1}$. 
This means that for lower $\dot{M}$ the propeller phase never starts, 
and the neutron star promptly starts life as a radio pulsar. 
At the other extreme extended exposure to the highest mass inflow rates may 
lead to accumulation of enough circumstellar material such that the optical 
thickness to electron scattering washes away any beaming of the X-rays 
from the neutron star and the rotation period is not detected 
(Lamb et al 1985, Ghosh, Angelini \& White 1997).

If all supernovae left neutron stars that
went through an AXP phase under mass inflow from the debris, then we would
expect at least 100 such objects for a lifetime $>$ 10$^4$ yrs and galactic
supernova rate of 10$^{-2}$, since the AXPs would be observable from all
galactic distances. That we only observe a few indicates that 
such sources rarely reach the accretion phase.

The number of DTNs must be much larger, since we see 
several within a distance of the order of 100 pc. In the 
Galaxy we would expect about
10$^4$ DTNs, and a lifetime of the order of 10$^6$ yrs or longer, for a
birth rate of 10$^{-2}$ yr$^{-1}$ or lower. The much smaller number of the
AXPs may be due to the depletion of the supply of mass inflow in most
sources by the end of the propeller phase. If the mass inflow is depleted
before the end of the propeller phase, subsequent accretion phases never 
occur. This may be the reason why the numbers of AXPs and SGRs are much less 
than the number of DTNs. By the same token, mass inflow 
rates $<$ 3.2 $\times$ 10$^{14}$ gm s$^{-1}$ are not inferred: at lower 
mass inflow rates the duration of the propeller phase exceeds the
lifetime of the mass inflow source. The rare AXPs and SGRs are thus the
young objects born with conventional magnetic fields but in circumstances of
large enough mass inflow towards the neutron star, so that the time t$_0$,
which scales with $\dot{M}^{-1}$, is short compared to the lifetime of the
mass supply: These are the rare sources which have gone through the
propeller phase in time to start accretion before the matter supply is
depleted. The more common source is the DTN, in a propeller phase of
duration 10$^6$ yrs, surviving the disappearance of the supernova
remnant. While it is reasonable to expect that the lifetime of the mass
supply increases at the lower mass inflow rates, one needs a model of
the disk evolution to make the comparison with the propeller spindown
quantitative. If DTNs are born at a rate comparable to the total
supernova rate, then a significant fraction of SNRs must
leave the neutron star under conditions of low enough 
$\dot{M}$. If all AXPs are to be descendants of DTNs, either the
propeller phase lasts 1000 times longer than the AXP accretion phase, or
only 10$^{-3}$ of DTNs turn into AXPs, according to the comparison of total
numbers of these two types of sources in the Galaxy. 
Taking into account the evidence for possible association with
supernova remnants, and the agreement between SNR ages and the
age estimates the asymptotic model, mass supply from a debris disk around a
single neutron star seems to be the likely scenario, rather than mass supply
from a binary companion.

The SGRs are like the AXPs in their X-ray properties. Both classes of
sources are detectable throughout the galaxy, and their comparable numbers
suggest the SGRs are also in the same rare or relatively short beginning
accretor phase as the AXPs. The claimed associations of most SGRs with 
SNRs are less certain than the AXP-SNR associations. Unlike the
AXPs the SGRs are located at the outskirts of their candidate SNR
associations. Nevertheless these SNRs are of the same atypical dense
class as the SNRs associated with the AXPs (Marsden et al 1999).

For the SGRs the magnetar hypothesis is on a different, stronger footing
than is the case for the AXPs, in that it provides the energy store and 
a detailed dyamical
model for the soft gamma ray bursts. While the present work does not address
the mechanisms of the gamma ray burst phenomenon, it is nevertheless
intriguing that as Wang and Robertson (1985) have noted propellers (and
probably their descendants the early accretors) can support relativistic
particle luminosities and gamma ray production in the surrounding
accumulated matter. The energy source could be sporadic accretion of
accumulated circumstellar mass released through instabilities.
The remnant disk scenarios were first recalled in connection with the
problem of planet formation around pulsars (Lin, Woosley \& Bodenheimer 1991), 
and disk instabilities
involving planet-like masses (Mineshige, Nomoto \& Shigeyama 1993) 
are consistent with the luminosities of the
SGRs. It may not be unreasonable to speculate that SGRs are also a
class of neutron stars with conventional fields undergoing disk
instabilities. 

Why is it that the AXPs are observed as pulsars while the LMXBs are not ?
The important single exception is the recently
discovered source XTE J1808-369. The explanation for the
rarity of LMXB pulsars has been that comptonization by circumstellar material 
will destroy pulses by washing out the beams emerging from the
neutron star if the material is optically thick to electron scattering 
(Lamb et al. 1985). This was noted in connection with the AXPs by Ghosh, 
Angelini \& White (1997).
The present picture is consistent with this: the AXPs are
observed as pulsars because they are in the beginning stages of accretion,
and the corona around them does not have significant optical 
thickness $\tau_{es} > $ 1 to destroy the beaming.
Work in progress to test this hypothesis by modelling the AXP spectra as
unsaturated comptonization and estimating the optical thicknesses will
be reported separately.

The radio quiet neutron stars found in centers of the supernova remnants
Cas A, Puppis A, RCW 103 and 296.5+10 (Chakrabarty et al. 2000, 
Gaensler, Bock \& Stappers 2000, Petre, Becker \& Winkler 1996, 
Gotthelf, Petre \& Hwang, 1997, Gotthelf, Petre \& Vasisht 1999, 
Mereghetti, Bignami \& Caraveo 1996, 
Vasisht et al. 1997) may be the propeller 
sources for which the cumulative effect of the mass inflow has indeed
set up a corona of $\tau_{es} > $ 1 around the neutron star. The source 
properties are very similar. Variability of the flux 
(Gotthelf, Petre \& Vasisht 1999) 
favours accretion or propeller driven energy dissipation rather than the 
cooling luminosity of an isolated neutron star. The low
luminosities, L$_x \sim$ 10$^{33}$ - 10$^{34}$ erg s$^{-1}$ suggest that 
these sources are in the propeller phase. Accretion would imply 
$\dot{M} \sim$ 10$^{13}$ - 10$^{14}$ gm/s, which can accrete onto the 
neutron star only if the rotation period is longer than 50 s (Eq.(6)) for 
B $\sim$ 10 $^{12}$ G. But then at these low accretion 
rates there is no reason that we should not see 
pulses at the rotation period. Blackbody effective 
areas of 1 km$^2$ also enhance the expectation of coherent pulsations at the 
rotation period from these sources while upper limits of 13\% and 25\% have 
been obtained for the RQNSs in RCW103 and Cas A respectively 
(Gotthelf, Petre \& Vasisht 1999, Chakrabarty et al. 2000). Taking the RQNSs 
to be propeller sources with L$_x$ = $\dot{E}_{diss}$, as  
we did for the DTNs, leads to $|\dot{\Omega}|\sim $10$^{-10}$ - 10$^{-8}$ 
rad s$^{-2}$. Since the rotation periods of the RQNSs are not known, we use 
Eq.(10) with B $\sim$ 10 $^12$ G, and obtain 
\begin{equation}
\dot{M} > 2 \times 10^{17} gm s^{-1}.
\end{equation}
For the RQNSs the mass inflow rates $\dot{M}$ are
indeed higher than those inferred for the DTNs and AXPs, consistent with  
the proposal that the radio quiet neutron stars are the 
high $\dot{M}$ end of our spectrum of neutron stars under mass inflow from 
remnant disks. Model parameters for the RQNS in Cas A are 
displayed in Table 1. The luminosity is typical of the RQNSs. As a propeller 
with luminosity arising from energy dissipation, taking 
$|\dot{\Omega}| > $ 3 $ \times $ 10$^{-10}$ rad s$^{-2}$ with $\mu_{30}$ = 1 
and obtain $\dot{M} >$ 7.4 $\times $ 10$^{17}$ gm s$^{-1}$, in agreement 
with the interpretation here. P$_{eq} < $1 s , corresponding to the high 
mass inflow rate.  The unobserved periods of the RQNSs are smaller than the 
observed range of periods of the AXPs and DTNs. 

\section{Luminosity of the Disk}

For both the DTNs and the AXPs there will be another source of energy 
dissipation in the disk or circumstellar material, due to accretion down 
to $r_A$: 
\begin{equation}
L(r_A) = 3/2 GM {\dot{M}} / r_A \sim 3/2 (GM)^{2/3}\Omega^{2/3} {\dot{M}}
\sim 1.3 \times 10^{33} erg s^{-1} {\dot{M}}_{15} m^{2/3} P^{-2/3}
\end{equation}
Near rotational equilibrium, L(r$_A$) = I$\Omega \dot{\Omega}$, the
power expended on the star by the spindown torque. 
For the AXPs this luminosity is smaller than the accretion luminosity by a
factor 
\begin{equation}
L(r_A)/L \sim 3/2 R/r_A (\dot{M}/\dot{M}_{acc}) \sim 10^{-2} 
(\dot{M}/\dot{M}_{acc}) m^{-1/3} P^{-2/3}
\end{equation}
noting that the mass inflow may not be all accreted even in the AXPs. For
the DTNs L(r$_A$) is actually larger than the observed 
luminosity $\dot{E}_{diss}$ by a factor 
\begin{equation}
L(r_A)/\dot{E}_{diss} \sim (I \Omega |\dot\Omega|)/ (I_p \omega
|\dot\Omega|) \sim (10^{2}-10^{4}) \Omega.
\end{equation}
If this luminosity is dissipated entirely at the boundary 
region, r $\sim$ r$_A$, the effective temperature is: 
\begin{equation}
T(r_A) = (L(r_A)/4 \pi {r_A}^2 \sigma)^{1/4} \sim 9.5 \times 10^4 K
{{\dot{M}
}_{15}}^{1/4} P^{-1/2}
\end{equation}
If the disk temperature lies in the euv range this would be extremely difficult
to detect, but being nearby sources, DTNs with low neutral hydrogen column
density might provide a chance of looking for a disk luminosity as a test of
the present model. For the thin disk model employed by Chatterjee, 
Hernquist \& Narayan (1999) 
calculations of the expected disk spectrum and lumionosity are reported by 
Perna, Hernquist \& Narayan (2000). 
For T(r$_A$) in the optical or IR, observational upper limits
for disks around the AXPs (Coe and Pightling 1998, Hulleman et al 1999)
pose a problem for thin disk models. The upper limits allow thin disks
with disk outer radii $\sim$ 10$^{10}$ cm, which requires a very young
age, t $\sim$ 100 yrs, in terms of the viscous evolution of 
an isolated thin disk. However several considerations suggest that the
remnant disks may be consistent with the observational upper bounds. 
The propeller effect might well lead to boundary conditions that alter the
disk evolution. The work done on the inflowing material by the
propeller, at the rate I$\Omega \dot{\Omega}$, may lift and sustain the
material at distances much larger than r$_A$. The emerging radiation may
have a lower effective temperature, from a larger effective area. 
The remnant disk may be a thick disk, and it may be enshrouded 
in a comptonizing corona, as we invoked above.  Modelling of remnant
disks and circumstellar material with propeller boundary conditions, and
future observations on the spectra of DTNs, AXPs and radio quiet
neutron stars will help to resolve these issues.

\section{Discussion and Conclusions}
 
A unified picture is proposed to account for all neutron stars formed in 
supernovae, and to include RQNSs, DTNs and AXPs and perhaps SGRs. 
The salient features of this picture are:
 
(i) The signature of a young neutron star depends on the presence and nature of 
the mass inflow of fallback material from the supernova explosion. 
The related classes of sources represent different pathways under 
different mass inflow rates and histories.
  
(ii) All neutron stars are born with 10$^{12}$ G fields.

(iii) Radio pulsars are formed if there is no mass inflow or not enough 
to protrude the light cylinder. Higher mass inflow rates do not allow 
radio pulsar activity, leading to DTNs, AXPs and RQNSs. 

(iv) The similarity in the rotation periods of these sources is not a 
coincidence but rather a
consequence of the asymptotic approach to rotational equilibrium under
a wide range of mass inflow rates.

(v) With low $\dot{M}$ the spindown is too slow for
the source to go through the propeller stage and reach the accretion stage
before the circumstellar mass is depleted. These sources, observed as DTNs,
are propellers for as long as 10$^5$-10$^6$ yrs, and
make up the most numerous class.

(vi) The DTNs are the first
observed examples of neutron stars in the propeller phase.

(vii) A non-accreting neutron star under propeller spindown has a luminosity
provided by energy dissipation inside the star.

(viii) The observability of the rotation period in the
AXPs but not in the LMXBs can be explained qualitatively in terms of
comptonization as supported by an interpretation of their spectra. The RQNSs 
are the young neutron stars for which comptonization washes out beaming so that the rotation period is not observable.

(vi) With low $\dot{M}$ the spindown is too slow for
the source to go through the propeller stage and reach the accretion stage
before the circumstellar mass is depleted. These sources, observed as DTNs, 
are propellers for as long as 10$^5$-10$^6$ yrs, and 
make up the most numerous class.

(vii) The DTNs are the first
observed examples of neutron stars in the propeller phase. 

(viii) A non-accreting neutron star under propeller spindown has a luminosity
provided by energy dissipation inside the star.

Arguments that AXPs and DTNs are magnetars are based on the grounds that
isolated neutron stars with ordinary 10$^{12}$ G fields cannot have spun
down to $\sim$ 10 s periods within the estimated ages. The above picture
obviates the need to postulate magnetars for AXPs and DTNs. For the SGRs the
required energy budgets and dynamical arguments make a case for the
magnetar hypothesis. Marsden et al (1999) have noted that AXPs and SGRs
share a common morphology of their associated supernova remnants, which 
they link with the high density of the surrounding interstellar medium. These
authors argue that the common properties of these classes of sources must
therefore be due to a common set of environmental factors, 
rather than intrinsic neutron star properties like superstrong
magnetic fields. Whether the large spindown rates observed are due to
spindown by interaction with ambient matter, as proposed here, as well as 
by Chatterjee, Hernquist \& Narayan (1999) and Marsden et al. (1999) or to
spindown by a magnetar can be decided by detailed analysis of the
fluctuations (noise) in the spindown process, as the timing noise
characteristic of accretion powered neutron stars is quite distinguishable
from the timing noise in the typically much quieter isolated rotation
powered pulsars. This analysis will require frequently sampled timing
observations of the AXPs and SGRs. The recently reported extended quiet
spindown phases in 1E2259+586 and 1RXS J170849.0-400910 
(Kaspi, Chakrabarty \& Steinberger 1999) 
is not sufficient to conclude that this source
is undergoing spindown under magnetic dipole radiation since the quiet
spindown phase was preceded by episodes of higher spindown rates with large
timing noise strengths (Baykal \& Swank 1996). Whether such changes in the 
torque repeat periodically as foreseen for a precessing magnetar (Melatos 1999) remains to be
checked in future timing observations. A well known accreting source,
the LMXB 4U1626-67, has exhibited similar quiet episodes as
well as intervals of strong timing fluctuations that are typical of
accreting sources (Chakrabarty et al. 1997). It is encouraging from the unified point of view of
this work that during the quiet spindown epoch that would allow the
dtection of glitches, 1E 2259+586 has indeed exhibited a glitch that is
very similar to the glitches of the Vela pulsar and other radio pulsars
(Kaspi, Lackey \& Chakrabarty 2000). 
If the AXPs are confirmed as accreting sources, the
glitch from 1E2259+586 will constitute strong support for our starting
hypothesis that all neutron stars have the same internal dynamics and
the associated energy dissipation rates. Observation of a spindown rate in the
expected range from RXJ0720.4-3125 or periods and spindown rates from
other DTNs would constitute strong evidence for the
propeller hypothesis. 

The framework proposed here leads to a program of related 
issues to be reported 
in subsequent work. These include models of the disk and circumstellar 
material, the emerging spectrum and pulse content, the time evolution 
of the mass inflow and the resulting spindown.

\acknowledgments

I thank H. \"Ogelman, \c{S}.Balman, A.Baykal, A.Esendemir, O. Guseinov, 
\"U.K{\i}z{\i}lo\u{g}lu and members of the High Energy Astrophysics 
Research Unit at METU for useful comments and \c{C}. \.{I}nam for help with 
the table and the figure. I acknowledge partial support from the Scientific 
and Technical Research Council of Turkey (T\"UB\.{I}TAK), from the Turkish 
Academy of Sciences and from the NSF International Grant 9417296
at the University of Wisconsin-Madison. 

\newpage \noindent  REFERENCES

\noindent 
Alpar,M.A. 1998, in Neutron Stars and Pulsars, N. Shibazaki, N. Kawai, 
S. Shibata \& T. Kifune, eds., (Universal Academic 
Publishers: Tokyo), 129 \newline
Alpar, M.A. 1999, http://xxx.lanl.gov/abs/astro-ph/9912228\newline
Alpar,M.A. \& Baykal,A. 1994, MNRAS, 269, 849\newline
Alpar,M.A. et al. 1984, ApJ, 276, 325\newline
Alpar,M.A. et al. 1987, A\&A, 276, 101\newline
Alpar,M.A., Nandkumar,R. \& Pines,D. 1985, 288, 191\newline
Alpar,M.A., Chau,H.F., Cheng,K.S. \& Pines,D. 1993, ApJ, 409, 345\newline
Baykal,A. \& Swank,J.H. 1996, ApJ, 460, 470\newline
Becker,W. \& Tr\"umper,J. 1997, A\&A, 326, 682\newline
Brazier, K.T.S. \& Johnston,S. 1999, MNRAS, 305, 671\newline
Campana,S. et al. 1998, ApJ, 499, L65\newline
Canizzo,J.K., Lee,H.M. \& Goodman,J. 1990, ApJ, 351,38\newline
Chakrabarty,D. et al. 1997, ApJ, 474, 414\newline
Chakrabarty,D. et al. 2000, http://xxx.lanl.gov/abs/astro-ph/0001026\newline
Chatterjee,P., Hernquist, L. \& Narayan,R. 1999, 
http://xxx.lanl.gov/abs/astro-ph/9912137\newline
Chevalier,R.A. 1989, ApJ, 346, 847\newline
Cline,T. et al. 1980, ApJ, 237, L1\newline
Coe,M.J. \& Pightling,S.L. 1998, MNRAS, 299,233\newline
Cui,W. et al. 1998, ApJ, 502, L49\newline
Duncan,R.C. \& Thompson,C. 1992, ApJ 392, L9\newline
Frank,J., King,A. \& Raine,D. 1992, Accretion Power in Astrophysics,
Cambridge University Press\newline
Gaensler,B.M. 1999, http://xxx.lanl.gov/abs/astro-ph/9911190\newline
Gaensler,B.M., Bock,D.C.-J. \& Stappers,B.W. 2000, 
http://xxx.lanl.gov/abs/astro-ph/0003032\newline
Gotthelf,E.V., Petre,R. \& Hwang,U. 1997, ApJ, 487, L175\newline
Gotthelf,E.V., Petre,R. \& Vasisht,G. 1999, 
http://xxx.lanl.gov/abs/astro-ph/9901371\newline
Ghosh,P. \& Lamb, F.K. 1991, in Neutron Stars: Theory and Observation, 
J. Ventura \& D. Pines, eds. (Kluwer:Dordrecht), 363\newline
Ghosh,P., Angelini,L.\& White,N.E. 1997, ApJ, 478, 713\newline
Haberl, F. et al. 1997, A\&A, 326, 662\newline
Haberl, F., Motch, C. \& Pietsch, W. 1998, Astron. Nachr., 319, 97\newline
Haberl, F., Pietsch, W. \& Motch, C. 1999, 
http://xxx.lanl.gov/abs/astro-ph/9911159\newline
Heyl, J.S. \& Hernquist, L. 1998, MNRAS 297, L69\newline
Hulleman, F. et al. 2000, http://xxx.lanl.gov/abs/astro-ph/0002474\newline
Hurley, K. et al. 1999, ApJ 510, L111\newline
Illarionov,A.F. \& Sunyaev,R.A. 1975, A\&A, 39, 185\newline
Kaspi, V.M. 1999, http://xxx.lanl.gov/abs/astro-ph/9912284\newline
Kaspi, V.M., Chakrabarty, D. \& Steinberger, J. 1999, 
http://xxx.lanl.gov/abs/astro-ph/9909283\newline
Kaspi, V.M., Lackey, J.R. \& Chakrabarty, D. 2000, ApJ, submitted\newline
Kouveliotou, C. et al. 1998, Nature 393, 235\newline
Kouveliotou, C. et al. 1999, ApJ 510, L115\newline
Kulkarni, S.R. \& van Kerkwijk, M.H. 1998, ApJ, 507, L49\newline
Lamb, F.K., Shibazaki, N., Alpar, M.A. \& Shaham, J. 1985, 
Nature, 317, 681\newline
Lin, D.N.C., Woosley, S.E. \& Bodenheimer, P.H. 1991, Nature, 353, 827\newline
Lipunov,V.M. 1992, Astrophysics of 
Neutron Stars, (Springer)\newline Marsden et al. 1999\newline
Marsden, D., Lingenfelter, R.E., Rothschild, R.E. \& Higdon, J.C. 1999, 
http://xxx.lanl.gov/abs/astro-ph/9912207\newline
Mazets, E.P. et al. 1979, Nature 282, 587\newline
Mazets, E.P. et al 1981, Ap\&SS 80, 3\newline
Melatos,A. 1999, ApJ, 519, L77\newline
Menou et al. 1999, ApJ, 520, 276\newline
Mereghetti, S. 1995, ApJ, 442, L17\newline
Mereghetti, S. 1999, http://xxx.lanl.gov/abs/astro-ph/9911252\newline
Mereghetti, S. \& Stella, L. 1995, ApJ, 442, L17\newline
Mereghetti, S., Bignami, G.F. \& Caraveo, P.A. 1996, ApJ, 464, 842\newline
Mereghetti, S., Israel, G.L. \& Stella, L. 1998, MNRAS, 296, 698\newline
Mineshige,S., Nomoto,K. \& Shigeyama,T. 1993, A\&A, 267, 95\newline
Motch, C. \& Haberl, F. 1998, A\&A, 333, 
L59\newline
Motch, C. et al. 1999, A\&A, in the press, astro-ph/9907306\newline
\"Ogelman,H. 1995, in The Lives of the Neutron Stars, 
M.A. Alpar, \"U. K{\i}z{\i}lo\u{g}lu \& J. van Paradijs eds., 
(Kluwer:Dordrecht), 101 \newline
Perna,R., Hernquist,L. \& Narayan, R. 2000, ApJ, submitted\newline
Petre, R., Becker, C.M. \& Winkler, P.F. 1996, ApJ, 465, L43\newline
Schwope et al. 1999,  A\&A, 341, L51\newline
Shemar,S.L. \& Lyne,A.G. 1996, MNRAS, 282, 677\newline
Stocke, J.T. et al. 1995, AJ, 109, 1199\newline
Thompson, C. \& Duncan, R.C. 1993, ApJ 408, 194\newline
Thompson, C. \& Duncan, R.C. 1995, MNRAS 275, 255\newline
Torii, K. et al. 1998, ApJ, 503, 843\newline
Treves et al. 1999, http://xxx.lanl.gov/abs/astro-ph/9911430\newline
Umeda,H. et al. 1993, ApJ, 408, 186\newline
van Paradijs,J.,Taam, R.E.\& van den Heuvel,E.P.J. 1995, A\&A 299, L41\newline
Vasisht, G. et al. 1997, ApJ, 476, L43\newline
Walter,F.M.,Wolk,S.J.\& Neuhauser,R. 1996, Nature 379, 233\newline
Walter,F.M.\&Matthews,L.D. 1997, Nature 389, 358\newline
Wang,Y.-M. \& Robertson,J.A. 1985, A\&A, 151, 361\newline
Wilson, C.A. et al. 1998, ApJ 513, 464\newline
Woods, P.M. et al. 1999, ApJ 519, L139\newline
Xi,X.-D. 1999, ApJ, 520, 271\newline
Yancopoulos,S., Hamilton,T.T. \& Helfand,D.J. 1994, ApJ, 429, 832\newline
Zhang,S.N., Yu,W. \& Zhang,W. 1998, ApJ, 494, L71\newline

\clearpage
\begin{figure}
\vspace{10cm}
\includegraphics{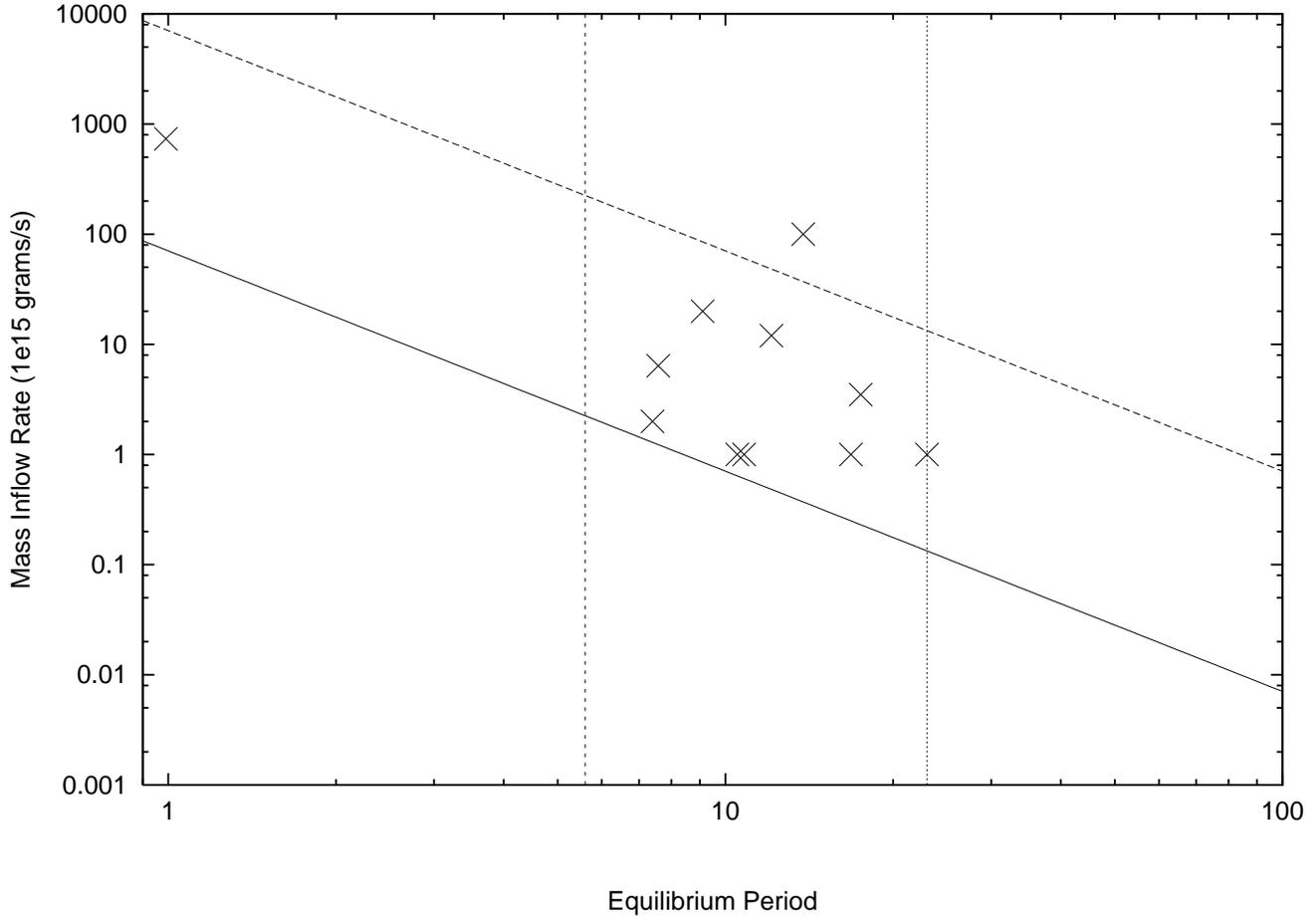}
\caption{Equilibrium periods, in seconds, and mass inflow rates,
in 10$^{15}$ gm s$^{-1}$, of the sources
in Table 1. 
The curves mark $\mu_{30}$ = 5 (upper) and
$\mu_{30}$ = 0.5 (lower). The two vertical lines bracket the range of observed
periods. Note the wide range of mass inflow rates.}
\end{figure}
\newpage
\vspace{0.5cm}
\begin{table}
\begin{center}
\begin{tabular}{clllllll}
\hline
Source & P (s) & $\dot{P}$ (s s$^{-1}$) & $\dot{M}_{15}$ & P$_{eq}$ (s) &
t$_o$ (yrs)
 & $\mu_{30}$  & t$_{SNR}$ (yrs) \\
 & & $\dot{\Omega}$ (rad s$^{-2}$) & & & & & \\
\hline
1E1048.1-5937 & 6.44 & 2 $\times$ 10$^{-11}$ & 1 & 23 & 7.4 $\times$ 10$^3$
 & 1.45  &  \\
 & & 0.2 & 64.4 & $9.3\times 10^3$ & 2.14 & \\
1E2259+586 & 6.98 & 6 $\times$ 10$^{-13}$ & 0.5 & 8.25 & $5.7\times 10^4$ & 0.67
 & \\
 & & & 2 & 7.4 & 1.7 $\times$ 10$^4$
 & 0.54 & $<2\times 10^4$ \\
           &      &                   & 100 & 6.99 & 367
 & 3.57 &  \\
4U0142+61 & 8.69 & 2.1 $\times$ 10$^{-12}$ & 1 & 10.5 & 2.1 $\times$ 10$^4$
 & 0.57 & \\
RXJ170849.0 & 11 & 2 $\times$ 10$^{-11}$ & 12 & 12.1 & 1.5
$\times$ 10$^3$ & 2.36 & \\
1E1841-045 & 11.8 & 4 $\times$ 10$^{-11}$ & 3.5 & 17.5 & 3.2 $\times$
10$^3$ & 1.96 & $ < 3 \times$ 10$^3$ \\
AXJ1845-0258 & 6.97 & ($7.8\times 10^{-12}$) & 1 & 16.8 & $1.2\times
10^{4}$ & (1) & $< 8 \times$10$^3$ \\
 & & (-$10^{-12}$) & & & & & \\
\hline
SGR1806-20 & 7.47 & 8.3 $\times$ 10$^{-11}$ & 100 & 13.8 & 148
 & 7.9 &  \\
SGR1900+14 & 5.16  & 1.1 $\times$ 10$^{-10}$ & 20 & 9.1 & 1.3 $\times$ 10$^3$
 & 2.2 &  \\
\hline
RXJ0720.4-3125 & 8.39 & (2.9 $\times$ 10$^{-11}$) & 1 & 10.8 & 2 $\times$
10$^4$
 & 0.59 & \\
 & & (-2.6$\times$ 10$^{-12}$) & & & & & \\
RXJ0420.0-5022 & 22.7 & $(8.2\times 10^{-10}$) & 6.4 & 7.58 & $5.27\times 10^{3}
$ & 3.6 & \\
 & & (-$10^{-11}$) & & & & & \\
\hline
RQNS in Cas A & & ($4.7\times 10^{-11}$) & 735 & 0.99 & 689 & (1) & \\
 & & (-3$\times$ 10$^{-10}$) & & & & & \\
\hline
\end{tabular}

\vspace{0.5cm}
\caption{Model parameters for all six AXPs, for the two 
SGRs with observed $P$ and $\dot{P}$, the two DTNs with observed periods, 
and the source in Cas A as a representative RQNS. 
For the AXPs and SGRs, mass inflow rates chosen
are greater than the mass accretion rate inferred from the observed
luminosity. For AXJ1845-0258, for which $\dot{P}$ has not been measured,
we assume $\mu_{30}=1$. For the two DTNs and for the RQNS in Cas A, we infer
$|\dot{\Omega}|$ from the luminosity. For the DTNs, we assume $P=P_{eq}$,
and for the Cas A source $\mu_{30}=1$ is assumed. Parentheses mark input quantities that are not taken from
observations.}

\end{center}
\end{table}
\end{document}